\def\fullversion{1}

\ifnum\fullversion=0
\documentclass[envcountsect,runningheads]{llncs}
\else
\documentclass{article}
\usepackage{fullpage}
\fi

\usepackage{times}
\usepackage{nicefrac}
\usepackage{cite}

\usepackage{comment}

\usepackage{amsmath, amsthm, amssymb, amsfonts}
\usepackage{graphicx}

\usepackage{prettyref}

\newcommand{\eat}[1]{}

\newcommand{\bbN}{\mathbb{N}}
\newcommand{\bbR}{\mathbb{R}}

\newcommand{\roundup}[1]{{\lceil {#1} \rceil}}
\newcommand{\rounddown}[1]{{\lfloor {#1} \rfloor}}

\DeclareMathOperator{\offline}{Offline}
\DeclareMathOperator{\online}{Online}

\newcommand{\eedf}{$e$-{\sc edf}}

\ifnum\fullversion=1

\newtheorem{theorem}{Theorem}[section]
\newtheorem{lemma}[theorem]{Lemma}
\newtheorem{proposition}[theorem]{Proposition}

\fi

\title{Online Algorithms for Machine Minimization}
\ifnum\fullversion=0
\author{Nikhil R. Devanur\inst{1} \and 
  Konstantin Makarychev\inst{1} \and
  Debmalya Panigrahi\inst{2} \and 
  Grigory Yaroslavtsev\inst{3}}
\institute{Microsoft Research; \{nikdev, komakary\}@microsoft.com \and
  Duke University, Durham, NC; debmalya@cs.duke.edu \and
  Brown University ICERM, Providence, RI; grigory@grigory.us}
\date{}
\authorrunning{Devanur, Makarychev, Panigrahi, Yaroslavtsev}
\else

\author{Nikhil R. Devanur
\\
Microsoft Research
\and
Konstantin Makarychev
\\
Microsoft Research
\and
Debmalya Panigrahi
\\
Duke University
\and
Grigory Yaroslavtsev
\\
Brown University, ICERM
}

\fi
\begin{document}

\maketitle

\begin{abstract}
In this paper, we consider the online version of the machine minimization problem 
(introduced by Chuzhoy~{\em et al}, FOCS 2004), where the goal is to schedule a 
set of jobs with release times, deadlines, and processing lengths on a minimum 
number of identical machines. Since the online problem has strong lower bounds
if all the job parameters are arbitrary, we focus on jobs with uniform length.
Our main result is a complete resolution of the deterministic complexity of this 
problem by showing that a competitive ratio of $e$ is achievable and optimal,
thereby improving upon existing lower and upper bounds of 2.09 and 5.2 respectively.
We also give a constant-competitive online algorithm for the case of 
uniform deadlines (but arbitrary job lengths); to the best of our knowledge, 
no such algorithm was known previously. Finally, we consider the complimentary
problem of throughput maximization where the goal is to maximize the sum of 
weights of scheduled jobs on a fixed set of identical machines (introduced by 
Bar-Noy {\em et al}, STOC 1999). We give a randomized online algorithm for this
problem with a competitive ratio of $\frac{e}{e-1}$; previous results achieved
this bound only for the case of a single machine or in the limit of an infinite 
number of machines.
\end{abstract}

%\newpage

\section{Introduction}
\label{sec:introduction}

Scheduling jobs on machines to meet deadlines is a fundamental area of
combinatorial optimization. In this paper, we consider a classical problem 
in this domain called {\em machine minimization}, which is defined as follows.
We are given a set of $n$ jobs, where job $j$ is characterized by a 
release time $r_j$, a deadline $d_j$, and a 
processing length $p_j$. The algorithm must schedule the jobs on a set of
identical machines such that each job is processed for a period $p_j$ 
in the interval $[r_j, d_j]$. The goal of the algorithm is to minimize
the total number of machines used. 

The machine minimization problem was previously considered in the offline 
setting, where all the jobs are known in advance. 
Garey and Johnson~\cite{GareyJ77, GareyJ79} showed that it is NP-hard to 
decide whether a given set of jobs can be scheduled on a single machine.
On the positive side, using a standard LP formulation and randomized 
rounding~\cite{RaghavanT87}, one can obtain 
an approximation factor of $O\left(\frac{\log n}{\log \log n}\right)$.
This was improved by Chuzhoy {\em et al}~\cite{ChuzhoyGKN04} to 
$O\left(\sqrt{\frac{\log n}{\log \log n}}\right)$ by using a more sophisticated LP formulation
and rounding procedure. This is the best approximation ratio currently known 
and it is an interesting open question as to whether a constant-factor 
approximation algorithm exists for this problem. (A constant-factor approximation 
was claimed in \cite{ChuzhoyC09a}, but the analysis is incorrect~\cite{ChuzhoyC09b}.)

In this paper, we consider the online version of this problem,
i.e., every job becomes visible to the algorithm when it is released. 
We evaluate our algorithm in terms of its 
{\em competitive ratio}~\cite{BorodinE98}, which is 
defined as the maximum ratio (over all instances) of the number of machines 
in the algorithmic solution to that in an (offline) optimal solution. For jobs 
with arbitrary processing lengths, an information-theoretic lower bound of 
$\Omega\left(\max(n, \log \left(\frac{p_{\max}}{p_{\min}}\right))\right)$
(where $p_{\max}$ and $p_{\min}$
are respectively the maximum and minimum processing lengths among all jobs)
was shown by Saha~\cite{Saha13}.\footnote{Saha~\cite{Saha13} only mentions 
the lower bound of 
$\Omega\left(\log \left(\frac{p_{\max}}{p_{\min}}\right)\right)$
but the same construction also gives a lower bound of $\Omega(n)$.} We note that 
matching (up to constants) upper bounds are easy to obtain. An upper
bound of $n$ in the competitive ratio is trivially obtained by scheduling
every job on a distinct machine. On the other hand, 
any algorithm with a competitive ratio of $\alpha$ for the case of
unit length jobs can be used as a black box to obtain an algorithm with a 
competitive ratio of $O(\alpha \log \frac{p_{\max}}{p_{\min}})$
for jobs with arbitrary processing lengths. 
%(For completeness, we show 
%this reduction in the appendix.)

%
%\begin{theorem}
%\label{thm:min-general-lower}
%	There is no deterministic algorithm for the online machine minimization
%	problem on $n$ jobs that has a competitive ratio of 
%	$o\left(\max(n, \log \left(\frac{p_{\max}}{p_{\min}}\right))\right)$.
%\end{theorem}

Our main focus in this paper is the online machine minimization problem 
with unit processing lengths. The previous best competitive 
ratio for this problem was due to Kao {\em et al}~\cite{KaoCRW12} who gave 
an upper bound of 5.2 and a lower bound of 2.09.
(Earlier, Shi and Ye~\cite{ShiY08} had claimed a competitive ratio of 2 for the special
case of unit job lengths and equal deadlines, but an error in their analysis
was discovered by Kao~{\em et al}~\cite{KaoCRW12}.)
Saha~\cite{Saha13} gave a different algorithm for this problem with a 
(larger) constant competitive ratio. Our main result in this paper 
is a complete resolution of the deterministic
online complexity of this problem by giving an algorithm that has a 
competitive ratio of $e$ and a matching lower bound.

\begin{theorem}
\label{thm:min-unit-optimal}
	There is a deterministic algorithm for the online machine minimization
	problem with uniform job lengths that has a competitive ratio of $e$. 
	Moreover, no deterministic algorithm for this problem has a competitive 
	ratio less than $e$. \footnote{It was brought to our attention that this theorem also follows from the results of~\cite{BKP07}, who consider energy-minimizing scheduling problems. This result follows from Lemma 4.7 and Lemma 4.8 in~\cite{BKP07}.}
\end{theorem}

%Our main technical result is Theorem~\ref{thm:min-unit-optimal} where we 
%show matching upper and lower bounds on the competitive ratio of the 
%online machine minimization problem with unit jobs.
%To obtain this result, we first restrict the set of algorithms.
%without changing the best competitive 
%ratio achievable. 
%Suppose $\alpha$ is the optimal competitive ratio.
To prove this theorem, we first show that the following 
online algorithm is the best possible: 
at any point of time, the algorithm schedules available jobs using  
{\em earliest deadline first} on $\alpha k$ machines, where $k$
is the number of machines in the optimal offline deterministic
solution for all jobs that have been released so far and $\alpha$
is the optimal competitive ratio. 
%We show that this algorithm is, in fact, optimal. 
Therefore, the main challenge is to
find the value of $\alpha$. If we under-estimate $\alpha$, then this
online algorithm is invalid in the sense that it will fail to schedule
all jobs. On the other hand, over-estimating $\alpha$ leads to a sub-optimal
competitive ratio. In order to estimate $\alpha$, we use an analysis technique 
that is reminiscent of factor-revealing LPs (see, e.g., Vazirani~\cite{Vazirani01}).
%(introduced by Jain and Vazirani~\cite{JainV01}). 
%We write an LP that 
%finds the worst possible strategy for releasing jobs.
%
%In this paper, we do not present the linear program as is. Instead, we give
However, instead of using the LP per se, we give
combinatorial interpretations of the primal and dual solutions.
We prove that the earliest deadline first schedule is valid if and only
if there exists a \emph{fractional schedule} satisfying two natural conditions
\emph{fractional completion} and \emph{fractional packing}. Then, we explicitly
construct an online fractional schedule (which corresponds to the optimal dual solution)
that uses the same number of machines as the online algorithm. To show a lower 
bound we explicitly present the offline strategy (which is essentially the optimal primal solution).

We also consider the online machine minimization problem where all the deadlines
are identical, but the processing lengths of jobs are arbitrary. For this 
problem, we give a deterministic algorithm with a constant competitive ratio.

\begin{theorem}
\label{thm:min-equal-deadline}
	There is a deterministic algorithm for the online machine minimization 
	problem with uniform deadlines that has a constant competitive ratio.
\end{theorem}

A problem that is closely related to the machine minimization problem 
is that of {\em throughput maximization}.
In this problem, every job $j$ has a given weight $w_j$ in addition to the 
parameters $r_j$, $p_j$, and $d_j$. The goal is now to schedule the maximum 
total weight of jobs on a given number of machines. If all the job parameters 
are arbitrary, then this problem has a lower bound of $\Omega(n)$ in the 
competitive ratio, even on a single machine where all jobs have unit weight. 
Initially, the adversary releases a job of processing length equal to its 
deadline (call it $d$) and depending on whether this job is scheduled or not, 
either releases $d$ unit length jobs with the same deadline, or does not 
release any job at all. As in the machine minimization problem,
we focus on the scenario where all jobs have uniform 
(wlog, unit) processing length. First, we note that in the unweighted case 
(i.e., all jobs have unit weight), it is optimal to use the 
{\em earliest deadline first} strategy, where jobs that are waiting to be
scheduled are ordered by increasing deadlines and assigned to the available 
machines in this order. Therefore, we focus on the weighted scenario.
Previously, the best online (randomized) algorithms for this problem 
had a competitive ratio of $\frac{e}{e-1}$ for the case of a single 
machine~\cite{BartalCCFJLST04, ChinCFJST06, Jez11}. 
For $k$ machines, Chin~{\em et al}~\cite{ChinCFJST06} obtained a 
competitive ratio of $\frac{1}{1 - \left(\frac{k}{k+1}\right)^k}$,
which is equal to 2 for $k = 1$ but decreases with increasing $k$
ultimately converging to $\frac{e}{e-1}$ as $k\rightarrow \infty$.
The best known lower bound for randomized 
algorithms is 1.25 due to Chin and Fung~\cite{ChinF03}, which holds
even for a single machine. 
%(For a discussion on
%the deterministic algorithms for this problem, the reader is referred to
%the section~\ref{sec:related}.) 

In this paper, we give an approximation-preserving reduction 
from the online throughput maximization problem with {\em any} number of 
machines to the online vertex-weighted bipartite matching problem.
(We will define this problem formally in section~\ref{sec:throughput}.)
%In the latter problem, the input comprises a bipartite graph 
%$(U \cup V, E)$, where the vertices in $U$ are given offline and 
%have non-negative weights associated with them. In each online step,
%a vertex in $V$ and its neighbors in $U$ are revealed and the algorithm
%must decide if it wants to match the new vertex with one of its
%neighbors in $U$ that has not been matched to any previous vertex 
%in $V$. The objective is to maximize the sum of weights of the 
%vertices in $U$ that are eventually matched with vertices in $V$.
Using this reduction and known algorithms for the online vertex-weighted
matching problem~\cite{AggarwalGKM11, DevanurJK13}, we obtain a 
randomized algorithm for the online throughput maximization problem
for unit length jobs that has a competitive ratio of $\frac{e}{e-1}$
{\em independent of the number of machines}.

\begin{theorem}
\label{thm:max-unit}
	There is a randomized algorithm for the online throughput maximization
	problem that has a competitive ratio of $\frac{e}{e-1}$, independent of the 
	number of machines.
\end{theorem}
%\enlargethispage*{10mm}
%\subsection{Our Techniques}
%\paragraph{\bf Our Techniques.}

%% Specifically, we encode the strategy of the online adversary as an LP 
%% and argue, using structural properties of the problem, that the optimal 
%% objective of the linear program is precisely equal to the value of $\alpha$.
%% Next, we construct the dual LP  for this program. The final step is to 
%% demonstrate feasible solutions for both the primal and dual LPs that have 
%% an objective value of $e$, which establishes the value of $\alpha$ to be $e$.

%\subsection{Related Work}
%\label{sec:related}

\noindent
{\bf Related Work.}
As mentioned previously, the offline version of the machine minimization 
problem was considered by Chuzhoy~{\em et al}~\cite{ChuzhoyGKN04} and 
Chuzhoy and Codenotti~\cite{ChuzhoyC09a}. 
Several special cases of this problem have also been considered. 
Cieliebak {\em et al}~\cite{CieliebakEHWW04} studied the problem under 
the restriction that the length of the time interval during which a job 
can be scheduled is small. Yu and Zhang~\cite{YuZ09} gave constant-factor
approximation algorithm for two special 
cases where all jobs have equal release times or equal processing lengths.

A related problem is that of scheduling jobs on identical machines where 
each job has to be scheduled in one among a given set of discrete intervals.
For this problem, an approximation hardness of $\Omega(\log \log n)$ was 
shown by Chuzhoy and Naor~\cite{ChuzhoyN04}, even when the optimal solution 
uses just one machine. The best approximation algorithm known for this problem
uses randomized rounding~\cite{RaghavanT87} and has an approximation factor 
of $O\left(\frac{\log n}{\log \log n}\right)$.

%As noted earlier, the online machine minimization problem was considered 
%earlier by Kao {\em et al}~\cite{KaoCRW12} and Saha~\cite{Saha13}. 

The complimentary problem of throughput maximization has a rich history
in the offline model. For arbitrary job lengths, the best known approximation 
ratios for the unweighted and weighted cases are respectively 
$\frac{e}{e-1}$~\cite{ChuzhoyOR06} and 
2~\cite{Bar-NoyGNS01, BermanD00}. On the other hand, the 
discrete version of this problem was shown to be MAX-SNP hard by 
Spieksma~\cite{Spieksma98}. Several variants of this problem have also been 
explored. E.g., when a machine is allowed to be simultaneously used by multiple jobs, 
Bar-Noy {\em et al}~\cite{Bar-NoyBFNS01} obtained approximation
factors of 5 and $\frac{2e-1}{e-1}$ for the weighted and unweighted
cases respectively. In the special case of every job having a single interval, 
Calinescu {\em et al}~\cite{CalinescuCKR11} have a 2-approximation algorithm,
which was improved to a quasi-PTAS by Bansal~{\em et al}~\cite{BansalCES06}.

As mentioned above, the previous best randomized algorithms for the online 
throughput maximization problem with unit length jobs were due to
Chin {\em et al}~\cite{ChinCFJST06} and Jez~\cite{Jez11}. For the case of 
a single machine, 
Kesselman {\em et al}~\cite{KesselmanLMPSS04} gave a deterministic algorithm with 
a competitive ratio of 2, which was improved to  $\frac{64}{33} \simeq 1.939$ by 
Chrobak {\em et al}~\cite{ChrobakJST07}. This was further improved to
$2\sqrt{2} - 1 \simeq 1.828$ by Englert 
and Westermann~\cite{EnglertW12} and, in simultaneous work, to  
$\frac{6}{\sqrt{5} + 1} \simeq 1.854$ by Li {\em et al}~\cite{LiSS07}. 
On the other hand, Andelman {\em et al}~\cite{AndelmanMZ03}, 
Chin and Fung~\cite{ChinF03}, and Hajek~\cite{Hajek01} showed a lower bound of 
$\frac{\sqrt{5}+1}{2} \simeq 1.618$ on the competitive ratio of any deterministic 
algorithm for this problem.

\eat{
\paragraph{\bf Roadmap.}
In Section~\ref{sec:unit-optimal}, we show Theorem~\ref{thm:min-unit-optimal} by giving an optimal deterministic
algorithm for the online machine minimization problem with unit job lengths. In Section~\ref{sec:equal-deadline},
we show Theorem~\ref{thm:min-equal-deadline} by giving a constant-competitive algorithm for the 
online machine minimization problem with uniform deadlines. Finally, in Section~\ref{sec:throughput}, we
give an approximation-preserving reduction from the online throughput maximization problem with unit job
lengths to the vertex-weighted online matching problem, which leads to Theorem~\ref{thm:max-unit}.
}

\section{Optimal Deterministic Algorithm for Unit Jobs}
\label{sec:unit-optimal}

In this section we present a deterministic online algorithm with competive ratio $e$, proving one half of
\prettyref{thm:min-unit-optimal}. In the next section,
we will prove that this algorithm is optimal, completing the other half.
%\begin{theorem}\label{thm:optimum-deterministic}
%For the online machine-minimizing unit job scheduling problem with arbitrary release times and deadlines there exists a deterministic online algorithm, which achieves competitive ratio $e$.
%%Furthermore, for any $\epsilon > 0$ no deterministic online algorithm can achieve  achieve competitive ratio
%% greater than $e - \epsilon$.
%\end{theorem}

The main challenge for the online algorithm is to determine how many 
machines to open at a given time $t$. The scheduling policy is easy: since all
jobs are unit jobs, we can simply use the {\em earliest deadline first} policy. 
The policy is that if we have $m(t)$ available machines at time $t$, we should pick $m(t)$ 
released but not yet completed jobs with the earliest deadlines and schedule them on these $m(t)$ machines
(if the total number of available jobs is less than $m(t)$ we schedule all available jobs at time $t$).
We formally prove that this policy is optimal in Lemma~\ref{lemma:fractional}. 

Note that the number of machines used by the offline solution is a constant over time. Consequently,
it is really easy to solve the offline problem -- we just need to find the optimal $m$ using binary search and then verify that the earliest deadline first schedule is feasible. 
To state the online algorithm we need to introduce the following notation: Let $\offline(t)$
be the offline cost of the solution for jobs released in the time interval $[0,t]$. That is,
we consider the subset of all jobs $J(t)=\{j: r_j\leq t\}$, and let $\offline(t)$ be the cost of the
optimal offline schedule for $J(t)$. Note that the algorithm knows all jobs in $J(t)$ at time $t$,
and thus can compute $\offline(t)$. 

\medskip
\noindent
\textbf{Algorithm \eedf:} The online algorithm uses $\roundup{e\,\offline(t)}$ machines
at time $t$. It uses the earliest deadline first policy to schedule jobs.

%\medskip

\begin{lemma}\label{lemma:fractional}
Consider a set of jobs $J$. Suppose we have $m(t)\in \bbN$ machines
at time $t\in \{0,\dots, T\}$. The earliest deadline schedule is feasible if and only if for 
every $d\in \{0,\dots, T\}$, there exists a collection of functions, a fractional solution, $\{f_j:[0,T]\to \bbR^+\}$ satisfying the following conditions (note that $f_j$'s depend on $d$):
%\todo{According to the statement, $f_j$ depends on $d$ but the notation shows no dependence. This is confusing unless clarified.} 
\begin{enumerate}
\item (Fractional completion) Every job $j$ with $d_j\leq d$ 
is completed before time $d$ according to the fractional schedule, i.e.,
%\begin{align*}
$\int_{r_j}^d f_j(x) dx = 1$.
%\end{align*}
Note that the fractional solution is allowed to schedule a job $j$ 
past its deadline $d_j$ (but before $d$).

\item (Fractional packing) For every $t \in [0,T]$ the total number of machines used according to the fractional schedule 
is at most $m(\rounddown{t})$, i.e.,
%\begin{align*}
$\sum_{\substack{j\in J\\d_j\leq d}} f_j(t)\leq m(\rounddown{t})$.
%\end{align*}
\end{enumerate} 

%Moreover, if there exists a feasible schedule that uses $m(t)$ machines at time $t$, then 
%there exists a set of functions $\{f_j\}$ as above.
\end{lemma}
\begin{proof}
In one direction -- ``only if'' -- this lemma is trivial. If the earliest deadline first schedule is feasaible and uses $m(t)$ 
machines at time $t$, then we let $f_j(t)=1$, if job $j$ is scheduled at time $\rounddown{t}$; and 
$f_j(t)=0$, otherwise. It is easy to see that $f_j$'s satisfy conditions both fractional completion and fractional packing conditions.

We now assume existence of $f_j$s that satisfy the conditions above and show that 
in the earliest deadline first schedule, no job $j$ misses its deadline $d_j$.
Fix $j^*\in J$ and let $d^*=d_{j^*}$. Let $\{f_j\}$ be the fractional schedule for $d^*$. Notice that we 
can assume that $f_j(x)=0$ for $x\leq r_j$ and $x\geq d^*$ (by simply redefining $f_j$ to be 0 for $x\leq r_j$ and $x\geq d^*$). Let $J^*=\{j:d_j\leq d^*\}$. 
Denote by $S(t)$ the set of jobs scheduled by the algorithm at one 
of the first $t$ steps $0,\dots, t-1$. We let $S(0)=\varnothing$. 

\begin{proposition}\label{prop:fractional}
For every $t\in\{0,\dots, T\}$ the following invariant holds: 
\begin{equation}
|S(t)\cap J^*| \geq \sum_{j\in J^*} \int_0^t f_j(x) dx.
\end{equation}
\end{proposition}
\begin{proof}
We prove this proposition by induction on $t$. For $t=0$, the inequality trivially holds, because both sides are equal 
to $0$. Now, we assume that the inequality holds for $t$, and prove it for $t+1$. There are two cases.

%%Case 1. 
In the first case, $m(t)$ jobs from $J^*$ are scheduled at time $t$. Then we have:
\begin{align*}
&|S(t+1)\cap J^*| = |S(t)\cap J^*| + m(t) \ge 
%\sum_{j\in J^*} \int_0^t f_j(x) dx + m(t) = 
\sum_{j\in J^*} \int_0^t f_j(x) dx + \int_t^{t+1} m(\rounddown{x}) dx \\
&\ge \sum_{j\in J^*} \int_0^t f_j(x) dx + \int_t^{t+1} \sum_{j\in J^*}  f_j(x) dx 
= \sum_{j\in J^*} \int_0^{t + 1} f_j(x) dx, 
\end{align*}
where the first equality uses the fact that all machines are busy at time $t$, the second inequality follows by the inductive hypothesis and the fourth inequality uses the fractional packing condition.

%%Case 2.
In the second case, the number of jobs from $J^*$ scheduled at time $t$ is strictly less than $m(t)$.
This means that all jobs in $J^*$ released by time $t$ are scheduled no later than at time $t$ (note that jobs in $J^*$ 
have a priority over other jobs according to the earliest deadline policy). In other words,
$J^*\cap J(t)\subseteq S(t+1)$. Together with the fact that $S(t+1)\subseteq J(t)$  this implies that $J^*\cap S(t+1) = J^*\cap J(t)$.
On the other hand, for $j\notin J(t)$, $r_j\geq t+1$ and, hence, $\int_0^{t+1} f_j(x) dx = 0$. Putting this together,
\begin{align*}
|J^*\cap S(t + 1)| = |J^*\cap J(t)| = \sum_{j\in J^*\cap J(t)} \int_0^{t+1} f_j(x)  dx \geq   \sum_{j\in J^*}  \int_0^{t+1}  f_j(x) dx,
\end{align*}
where the second equality follows by the fractional completion condition.
This finishes the proof of the proposition.%\qed
\end{proof}

By Proposition~\ref{prop:fractional}, the number of jobs from $J^*$ scheduled by time $d^*$ is:
$$|J^* \cap S(d^*)| \ge \sum_{j\in J^*} \int_0^{d^*} f_j(x) dx = \sum_{j\in J^*} \int_{r_j}^{d^*} f_j(x) dx = |J^*|.$$
Here, we used the fractional completion condition $\int_{0}^{d^*} f_j(x) dx =1$. Thus, all jobs in $J^*$ are scheduled before the deadline $d^*$. Consequently, 
the job $j^*$ does not miss the deadline $d^*= d_j$.%\qed
\end{proof}

%\begin{proof}[Theorem \ref{thm:optimum-deterministic}]
We now use Lemma~\ref{lemma:fractional} to prove that the schedule produced 
by Algorithm~\eedf\ is feasible. Pick
an arbitrary deadline $d^*\in \{0,\dots, T\}$, and let $J^* =\{j:d_j\leq d^*\}$. We fractionally
schedule every job $j\in J^*$ in the time interval $[r_j, d^* - (d^*-r_j)/e]$. We let
$$f_j(x)=
%%\frac{\mathbf{1} (x\in [r_j, d^* - \frac{(d^*-r_j)}{e})}{d^*-r_j}\equiv
\begin{cases}
\frac{1}{d^* - x},&\text{if } x\in \big[r_j, d^* - \frac{(d^*-r_j)}{e}\big];\\
0, &\text{otherwise}.
\end{cases}
$$

Note that the fractional schedule depends on $d^*$, and possibly $d^* - (d^*-r_j)/e > d_j$ for some $j$. So the fractional 
solution may run a job $j$ even after its deadline $d_j$ is passed. Nevertheless, as we show now,
functions $f_j$ satisfy the conditions of Lemma~\ref{lemma:fractional}.

First, we check the fractional completion condition. For $j\in J^*$, we have:
$$\int_{r_j}^{d^*} f_j(x) dx = \int_{r_j}^{d^*-(d^*-r_j)/e} \frac{dx}{d^* - x} = \ln \frac{d^* - r_j}{(d^*-r_j)/e} = 1.$$

We now verify the fractional packing condition. We consider a fixed $t \in [0,T]$, 
and show that this condition holds for this $t$. 
Note that $f_j(t)\neq0$ if and only if $t \in [r_j, d^* - \frac{(d^*-r_j)}{e}]$, 
which is equivalent to 
$d^* - e(d^* - t) \leq r_j \leq t$. We denote $d^* - e(d^* - t)$ by $r^*$,
and let $\Lambda = \{j\in J^* |  r^* \leq r_j\leq t\}$.
Thus, $f_j(t)\neq 0$ if and only if $j\in \Lambda$.
Then,
\begin{equation}
\sum_{j\in J^*} f_j(t) = \sum_{j\in J^*} 
\frac{\mathbf{1} \big(t\in [r_j, d^* - \frac{(d^*-r_j)}{e}]\big)}{d^*-t} = \sum_{j\in \Lambda} \frac{1}{d^*-t} = \frac{|\Lambda|}{d^* - t}.
\label{eq:sumF}
\end{equation}

We need to compare the right hand side of~(\ref{eq:sumF}) with $m(t) \equiv \roundup{e \offline (t)}$. 
By the definition of $\offline(t)$, all jobs in $\Lambda\subset J(t)$ can be 
scheduled on at most $\offline(t)$ machines. All jobs in $\Lambda$ must be completed by time $d^*$ 
(since $\forall j\in \Lambda\subset J^*$, $d_j\leq d^*$).
The number of completed jobs is bounded by the number of available ``machine hours'' in the time interval $[r^*,d^*]$ which is $(d^*-r^*)\times\offline (t)$.
Therefore,
$|\Lambda|\leq (d^*-r^*)\times\offline (t)$, and, since $(d^* - r^*) =  e (d^* - t)$,
$$\sum_{j\in J^*} f_j(t)=  \frac{|\Lambda|}{d^* - t}\leq\frac{(d^*-r^*)\times\offline (t)}{d^* - t} = e\,\offline(t).$$
This concludes the proof that $f_j$ satisfy the fractional completion and packing conditions, and thus, by Lemma~\ref{lemma:fractional}, 
the online schedule is feasible i.e., every job $j$ is completed before its deadline $d_j$. 

At every point of time $t$, Algorithm~\eedf\ uses $\roundup{e\cdot \offline (t)}$ machines; $\offline (t)$ is a lower bound on the cost of the
offline schedule. Hence, the Algorithm~\eedf\ is $e$ competitive.
%\qed
%\end{proof}

\section{Optimal Deterministic Lower Bound}

We now prove the second part of Theorem~\ref{thm:min-unit-optimal}, giving a lower bound on the competitive ratio of a deterministic online algorithm. This bound holds even if all deadlines are the same. 
We present an adversarial strategy that forces any deterministic online algorithm to open $(e-\varepsilon)\times \offline$ machines. 
As in the previous section, we let $\offline (t)$ to be the offline optimum solution for jobs in $J(t)$ i.e. for jobs released in the time
interval $[0,t]$. Let $\online(t)$ be the number of machines used by the online algorithm at time $t$.

\medskip
\noindent
\textbf{Adversary:} We fix a sufficiently large number $n$ and $N=n^2$. At time $t\in \{0,\dots, T\}$, the adversary releases $\rounddown{N/(n-t)}$ unit jobs with deadline $n$.
The stopping time $T$ equals the first $\tau$ such that $\online (\tau)\geq e\,\offline (\tau)$ if such $\tau$ exists, and $n-1$ otherwise.

%\medskip

First we prove the following auxiliary statement.

\begin{lemma}\label{lem:offline-opt-ub}
For all $t^*\in [0,n]$ it holds that  $\offline(t^*)\leq \roundup{N/(e(n-t^*))}$. 
\end{lemma}
\begin{proof}
We need to show
that all jobs in $J(t^*)$ can be scheduled on $m(t^*) = \roundup{N/(e(n-t^*))}$ machines.
Since all jobs have the same deadline we shall use a greedy strategy: at every point of time we run arbitrary $m(t^*)$ jobs if there are $m(t^*)$ available jobs; and all available jobs otherwise.
In the offline schedule the number of machines equals $m(t^*)$ and does not change over time.
The number of jobs released by the adversary at time $t$ increases as a function of $t$. Thus,
till a certain integral point of time $s^*$, the number of available machines is greater than the number of available jobs, and thus all jobs are executed
immediately after they are released. After that point of time, $m(t^*)$ machines are completely loaded with jobs until all jobs are processed. The number of jobs released in the time interval $[s^*, t^*]$ is 
upper bounded by
$$\int_{s^*}^{t^*} \frac{N}{n-x} dx = N \,\ln\frac{n- s^*}{n-t^*}.$$
The number of jobs that can be processed in the time interval $[s^*, n]$ is equal to
$$m(t^*)\times (n-s^*) \geq \frac{N}{e(n-t^*)} \times (n-s^*).$$
Note that  
$$N \,\ln\frac{n - s^*}{(n-t^*)} \leq N\,\frac{n - s^*}{e(n-t^*)},$$
since for every $x$, particularly for $x= (n-s^*)/(n-t^*)$, $\ln x \leq x/e$ (the minimum of 
$x/e - \ln x$ is attained when $x=e$). Therefore, all jobs are completed till the deadline $n$. This completes the proof.
\qed
\end{proof}

We are now ready to prove the second part of Theorem~\ref{thm:min-unit-optimal}. Consider a run of a deterministic algorithm. We need to show that $\online (n)\geq (e -\varepsilon) \offline (n)$. If for some $\tau$,
$\online (\tau)\geq (e-\varepsilon) \,\offline (\tau)$, then we are done:
$\online (n)\geq \online (\tau) \geq (e-\varepsilon) \ \offline (\tau) = (e-\varepsilon) \ \offline (n)$ (we have $\offline (\tau) = \offline (n)$ 
since
the adversary stops releasing new jobs after time $\tau$). So we assume that $\online (t) <  (e-\varepsilon)\,\offline (t)$ for all $t\in \{0,\dots, n-1\}$.

By Lemma~\ref{lem:offline-opt-ub} $\offline(t)\leq \roundup{N/(e(n-t))}\leq N/(e(n-t)) + 1$. By the assumption $\online (t) < (e - \epsilon) \offline(t) = (1 - \varepsilon/e) N/(n-t) + O(1)$.
The total number of jobs processed by the online algorithm is upper bounded by
\begin{align}
& \sum_{t=0}^{n-1} \online(t) \leq \int_0^{n-1} \online (x) dx + \online(n-1)\notag \\
&\leq \big(1 - \frac{\varepsilon}{e}\big) \Big(\int_0^{n-1} \frac{N}{n - x} dx + N + O(n)\Big)
\leq \big(1 - \frac{\varepsilon}{e}\big) N\ln n + N + O(n).\label{eq:proc_jobs}
\end{align}
On the other hand, the total number of jobs released by the adversary is lower bounded by 
\begin{equation}
\int_{0}^{n-1} \Big(\frac{N}{n-x} - 1\Big)\,dx = N \,\ln n - (n-1).\label{eq:rel_jobs}
\end{equation}
We get a contradiction since for a sufficiently large $n$, expression (\ref{eq:rel_jobs}) is larger than 
(\ref{eq:proc_jobs}).

%\begin{comment}

\section{Online Machine Minimization with Equal Deadlines}
\label{sec:equal-deadline}

In this section, we give a 16-approximation algorithm 
for the Online Machine Minimization problem 
with arbitrary release times and job sizes, but equal 
deadlines (thus proving Theorem \ref{thm:min-equal-deadline}). 
 We assume w.l.o.g. that the common deadline $d = 2^k-1$. 

\medskip
\noindent
\textbf{Algorithm.} The algorithm splits the time line $[0,d-1]$ into $k$ phases: 
$[0,\nicefrac{1}{2}(d+1)], [\nicefrac{1}{2}(d+1), \nicefrac{3}{4}(d+1)], \dots$. The length 
of phase $i$ is $\ell_i = 2^{k-i} = (d+1)/2^i$; we denote the beginning of the phase by $a_i$ 
and the end of the phase by $b_i$. In any phase $i$, when a new job $j$ is released, 
the algorithm classifies it as a \emph{short} job if the size of the job 
$s_j\leq \ell_i/4$ and as a {\em long} job otherwise.
If $j$ is long, the algorithm opens a new machine and executes $j$ right away;
otherwise, the algorithm postpones the execution of job $j$ to the next phase. 

At the beginning of phase $i$, the algorithm closes all machines that are idle 
(closed machines can be reopened later if necessary). Next, it splits the open 
machines into two pools: those serving long jobs and those serving short jobs
according to the following rule. A machine serves long jobs if the remaining length 
of the job currently being processed on it is at least $\ell_i/4$; otherwise, 
the machine serves short jobs. Note that some machines in the short jobs pool 
are actually serving jobs that were long when they were released in a previous
phase, but have become short now based on their remaining length. 
Then, the algorithm schedules all postponed short jobs (that were released in 
phase $i-1$) on machines in the short jobs pool. (Note that machines
in the short jobs pool are currently serving long jobs from previous phases
that have now become short. When scheduling the postponed short jobs, the 
algorithm allows these running jobs to complete first.) The algorithm 
assigns postponed short jobs to machines using
an offline greedy algorithm: it picks a postponed job $j$ and schedules it on 
one of the available machines serving short jobs if that machine can process 
$j$ before the end of the current phase. If there are no such machines, the algorithm 
opens a new machine (or reactivates as idle machine) and adds it to the pool 
serving short jobs. Note that once all 
postponed jobs are scheduled in the time interval $[a_i, b_i]$, the algorithm 
does not assign any new job to the pool serving short jobs since all short jobs
released in phase $i$ will be postponed to phase $i+1$.

\medskip
\noindent
\textbf{Analysis.} We prove that at every point of time, 
the number of machines serving short jobs does not exceed $8\offline + 1$, 
and the number of machines serving long jobs does not exceed $8\offline$. 
(Recall that $\offline$ denotes the number of machines in an optimal 
offline solution.)
Our proof is by induction on the current phase $i$. 
Observe that at the end of phase $i$, all machines serving 
short jobs are going to be idle, because all jobs scheduled in this 
phase must be completed by $b_i$. These machines will be closed at 
the beginning of phase $(i+1)$. Thus, 
the only machines that may remain open when we transition from 
phase $i$ to phase $(i+1)$ 
are machines serving long jobs. The number of such machines is at most 
$8\offline$ by the inductive hypothesis. The next two lemmas show that 
this property implies that 
$M_{short}(i+1)$ (resp., $M_{long}(i+1)$) --- the number of machines 
serving short jobs (resp., long jobs) in phase $(i+1)$ --- is at most
$8\offline + 1$ (resp., $8\offline$).

\begin{lemma}
\label{lma:short}
If the number of active machines at the beginning of phase $(i+1)$ is
at most $8\offline$, then $M_{short}(i+1) \leq 8\offline + 1$.
\end{lemma}
\begin{proof}
At the beginning of phase $(i+1)$, we schedule all postponed jobs. 
There are two cases. The first case is that we schedule all postponed jobs on the 
machines that remained open from the previous phase. 
As we just argued, the number of such machines is at most $8\offline$. 
The second case is that we opened 
some extra machines. This means that every machine serving short jobs 
(except possibly the last one that we open) finishes processing jobs in phase $(i+1)$
no earlier than time $b_{i+1} - \ell_{i}/4 = b_{i+1} - \ell_{i+1}/2$. 
Otherwise, if some machine was idle at at time $b_{i+1} - \ell_{i}/4$,
we would assign an extra short job to this machine instead of opening 
a new machine (note that the length of any short job is at most $\ell_{i}/4$). 
Therefore, every machine (but one) is busy for at least half of the time 
in phase $(i+1)$. The volume of work they process is lower bounded 
by $(M_{short}(i+1) - 1)\times \ell_{i+1}/2$.
%where $M_{short}(i+1)$ is the number of machines serving short jobs in phase $(i+1)$.

We now need to find an upper bound on the volume of work. 
Every short job that was released in phase $i$ and got postponed to phase $(i+1)$  must be
scheduled by the offline optimal solution 
in the time interval $[a_i, d]$ 
(recall that $a_i$ is the beginning of phase $i$
and $d$ is the deadline). A job $j$ that was initially classified as a long job
on release but got reclassified as a short job in phase $(i+1)$ must also be partially 
scheduled by the optimal solution in the time interval $[a_i, d]$. 
For such a job, the optimal solution must schedule at least the same amount of work 
for the time interval $[a_{i+1},d]$ as the online algorithm does for the time 
interval $[a_{i+1}, b_{i+1}]$. This follows from the fact that the online algorithm 
executed job $j$ right after it was released (since it was initially a long job),
and thus, no matter how $j$ is scheduled in the optimal solution, 
the remaining size of $j$ at time $a_{i+1}$ in the optimal solution
must be at least that in the online algorithm. Thus, the amount of work done by machines serving 
short jobs in phase $(i+1)$ in the online schedule 
does not exceed the amount of work done 
by all machines in the time interval $[a_i, d]$ in the optimal schedule. The latter
quantity is upper bounded by $(d-a_i)\times \offline < 2\ell_i \times \offline$. 
Consequently, $M_{short}(i+1) \leq 8 \offline + 1$.
\end{proof}

\begin{lemma}
\label{lma:long}
If the number of active machines at the beginning of phase $(i+1)$ is
at most $8\offline$, then $M_{long}(i+1) \leq 8\offline$.
\end{lemma}
\begin{proof}
%We now give a bound on $M_{long}(i+1)$ -- the number of machines serving long jobs 
%in phase $(i+1)$. 
Each long job released in phase $(i+1)$ or the remainder of a job that got classified 
as long in phase $(i+1)$ must have size at least $\ell_{i+1}/4$. Therefore, the total 
volume of these jobs is $\ell_{i+1}/4 \times M_{long}(i+1)$. The same argument as
we used for short jobs shows that all this volume must be scheduled by the offline 
solution in the time interval
$[a_{i+1},d]$. Hence, $(\ell_{i+1}/4) \times M_{long}(i+1)\leq 2\ell_{i+1} \times \offline$ 
and therefore, $M_{long}(i+1)\leq 8\offline$. %This concludes the proof.
\end{proof}

\section{Online Throughput Maximization for Unit Jobs}
\label{sec:throughput}

In this section, we will give a reduction of the online throughput 
maximization problem for unit length jobs to the online 
vertex-weighted matching problem~\cite{AggarwalGKM11,DevanurJK13}. 
We will reuse the notation for the throughput maximization problem 
from the introduction, i.e., a job is characterized by it 
release time $r_j$, deadline $d_j$, and weight $w_j$.
%First, let us establish notation for both problems. In the throughput
%maximization problem, weighted unit length jobs $(r_j, d_j, w_j)$
%characterized by their release time $r_j$, deadline $d_j$, and 
%weight $w_j$ arrive online, i.e. they are revealed to the algorithm 
%at time $r_j$. As usual, we will assume that all release times
%and deadlines are integral. Each job can either be scheduled on
%a machine in any time step in the interval $[r_j, d_j]$ or not
%scheduled at all. Given a set of $k$ identical machines, each 
%of which can process at most one job in any time step, the goal 
%of the algorithm is to maximize the sum of weights of scheduled
%jobs, which we also call the throughput of the schedule.
Let us now define the online vertex-weighted matching problem.
The input comprises a bipartite graph
$G = (U\cup V, E)$, where the vertices in $U$ are given offline 
and have weights $w_u, u\in U$ associated with them and the 
vertices in $V$ (and their respective incident edges) appear
online. When a vertex in $V$ appears, it must be matched to one
of its neighbors in $U$ that has not been matched previously
or not matched at all. The goal of the algorithm is to maximize
the sum of weights of vertices in $U$ that are eventually 
matched by the algorithm. 
%The classical online matching
%problem is a special case of this problem when all the vertex 
%weights are uniform. For the vertex-weighted matching problem,
Aggarwal~{\em et al} \cite{AggarwalGKM11} introduced this problem
%generalized the 
%celebrated {\sc Ranking} algorithm of 
%Karp, Vazirani, and Vazirani~\cite{KarpVV90} to obtain a 
and obtained a randomized algorithm with a competitive ratio of 
$\frac{e}{e-1}$.
An alternative proof of this result was recently obtained by 
Devanur, Jain, and Kleinberg~\cite{DevanurJK13} using a 
randomized version of the classical primal dual paradigm.

Our main contribution is an approximation preserving reduction 
from the online throughout maximization problem to the 
online vertex-weighted matching problem.
%, which allows us to use the existing
%algorithms for the matching problem and obtain a randomized 
%competitive ratio of $\frac{e}{e-1}$ for throughput 
%maximization, thereby proving Theorem~\ref{thm:max-unit}.
Suppose we are given an instance of the throughput maximization
problem. Define an instance of the vertex-weighted matching problem
as follows. For every job $j$, define an offline vertex $u_j\in U$
with weight $w_j$. 
For every time step $t$, define $k$ online vertices in $V$, one for 
each machine. Let $v_{it}$ denote the vertex for machine $i$
in time step $t$, and add edges between each such online vertex and all 
offline vertices representing jobs $j$ such that $t\in [r_j, d_j]$.

Note that the reduction is somewhat counter-intuitive in that online
jobs are being mapped to the offline side of the bipartite graph.
However, it does produce a valid instance of the online 
vertex-weighted matching problem since at time $r_j$, both the offline
vertex corresponding to job $j$ and its first set of online neighbors 
(all online vertices corresponding to time step $r_j$) are simultaneously
revealed. 
%However, before describing the reduction, we identify a 
%property of the online vertex-weighted matching problem 
%that will be crucial to us later. 
This is sufficient since every offline 
vertex in $U$ comes into play in {\em any} algorithm 
only {\em after} its first online neighbor in $V$ is revealed. 
%Therefore, it is not 
%required that the entire offline side of the bipartite graph 
%is available to the algorithm at the outset; the only property 
%required is that at any point of time, the weights of all 
%neighbors of the online vertices that have already appeared 
%are available to the algorithm.

We will now show that there is a 1-1 mapping between solutions of the 
throughput maximization instance and the vertex-weighted matching 
instance. Consider any solution to the matching instance. If an
offline vertex $u_j$ is matched to a neighbor $v_{it}$, then we 
schedule job $j$ at time $t$ on machine $i$. This is valid since 
the edge between $u_j$ and $v_{it}$ testifies to the fact that 
$t\in [r_j, d_j]$. Conversely, consider a solution for the throughput
maximization problem. If job $j$ is scheduled on machine $i$ at time
$t$, then add the edge between $u_j$ and $v_{it}$ to the matching.
First, note that this edge exists since  job $j$ could only have been 
scheduled at some time$t\in [r_j, d_j]$; and second,
the edges selected form a matching since no job is scheduled more than 
once and no machine can have more than one job scheduled on it in 
the same time step.

This completes the reduction and therefore, the proof of 
Theorem~\ref{thm:max-unit}.

%\input arbitrary-lower

%\newpage

%%\bibliographystyle{splncs} 
\bibliographystyle{plain} 
\bibliography{ref}

\begin{thebibliography}{10}

\bibitem{AggarwalGKM11}
Gagan Aggarwal, Gagan Goel, Chinmay Karande, and Aranyak Mehta.
\newblock Online vertex-weighted bipartite matching and single-bid budgeted
  allocations.
\newblock In {\em SODA}, pages 1253--1264, 2011.

\bibitem{AndelmanMZ03}
Nir Andelman, Yishay Mansour, and An~Zhu.
\newblock Competitive queueing policies for qos switches.
\newblock In {\em SODA}, pages 761--770, 2003.

\bibitem{BansalCES06}
Nikhil Bansal, Amit Chakrabarti, Amir Epstein, and Baruch Schieber.
\newblock A quasi-ptas for unsplittable flow on line graphs.
\newblock In {\em STOC}, pages 721--729, 2006.

\bibitem{BKP07}
Nikhil Bansal, Tracy Kimbrel, and Kirk Pruhs.
\newblock Speed scaling to manage energy and temperature.
\newblock {\em J. ACM}, 54(1), 2007.

\bibitem{Bar-NoyBFNS01}
Amotz Bar-Noy, Reuven Bar-Yehuda, Ari Freund, Joseph Naor, and Baruch Schieber.
\newblock A unified approach to approximating resource allocation and
  scheduling.
\newblock {\em J. ACM}, 48(5):1069--1090, 2001.

\bibitem{Bar-NoyGNS01}
Amotz Bar-Noy, Sudipto Guha, Joseph Naor, and Baruch Schieber.
\newblock Approximating the throughput of multiple machines in real-time
  scheduling.
\newblock {\em SIAM J. Comput.}, 31(2):331--352, 2001.

\bibitem{BartalCCFJLST04}
Yair Bartal, Francis Y.~L. Chin, Marek Chrobak, Stanley P.~Y. Fung, Wojciech
  Jawor, Ron Lavi, Jiri Sgall, and Tom{\'a}s Tich{\'y}.
\newblock Online competitive algorithms for maximizing weighted throughput of
  unit jobs.
\newblock In {\em STACS}, pages 187--198, 2004.

\bibitem{BermanD00}
Piotr Berman and Bhaskar DasGupta.
\newblock Improvements in throughout maximization for real-time scheduling.
\newblock In {\em STOC}, pages 680--687, 2000.

\bibitem{BorodinE98}
Allan Borodin and Ran El-Yaniv.
\newblock {\em Online Computation and Competitive Analysis}.
\newblock Cambridge University Press, New York, NY, USA, 1998.

\bibitem{CalinescuCKR11}
Gruia C{\u{a}}linescu, Amit Chakrabarti, Howard~J. Karloff, and Yuval Rabani.
\newblock An improved approximation algorithm for resource allocation.
\newblock {\em ACM Transactions on Algorithms}, 7(4), 2011.

\bibitem{ChinCFJST06}
Francis Y.~L. Chin, Marek Chrobak, Stanley P.~Y. Fung, Wojciech Jawor, Jiri
  Sgall, and Tom{\'a}s Tich{\'y}.
\newblock Online competitive algorithms for maximizing weighted throughput of
  unit jobs.
\newblock {\em J. Discrete Algorithms}, 4(2):255--276, 2006.

\bibitem{ChinF03}
Francis Y.~L. Chin and Stanley P.~Y. Fung.
\newblock Online scheduling with partial job values: Does timesharing or
  randomization help?
\newblock {\em Algorithmica}, 37(3):149--164, 2003.

\bibitem{ChrobakJST07}
Marek Chrobak, Wojciech Jawor, Jiri Sgall, and Tom{\'a}s Tich{\'y}.
\newblock Online scheduling of equal-length jobs: Randomization and restarts
  help.
\newblock {\em SIAM J. Comput.}, 36(6):1709--1728, 2007.

\bibitem{ChuzhoyC09b}
Julia Chuzhoy and Paolo Codenotti.
\newblock Erratum: Resource minimization job scheduling.
\newblock In {\em APPROX-RANDOM}, 2009.

\bibitem{ChuzhoyC09a}
Julia Chuzhoy and Paolo Codenotti.
\newblock Resource minimization job scheduling.
\newblock In {\em APPROX-RANDOM}, pages 70--83, 2009.

\bibitem{ChuzhoyGKN04}
Julia Chuzhoy, Sudipto Guha, Sanjeev Khanna, and Joseph Naor.
\newblock Machine minimization for scheduling jobs with interval constraints.
\newblock In {\em FOCS}, pages 81--90, 2004.

\bibitem{ChuzhoyN04}
Julia Chuzhoy and Joseph Naor.
\newblock New hardness results for congestion minimization and machine
  scheduling.
\newblock In {\em STOC}, pages 28--34, 2004.

\bibitem{ChuzhoyOR06}
Julia Chuzhoy, Rafail Ostrovsky, and Yuval Rabani.
\newblock Approximation algorithms for the job interval selection problem and
  related scheduling problems.
\newblock {\em Math. Oper. Res.}, 31(4):730--738, 2006.

\bibitem{CieliebakEHWW04}
Mark Cieliebak, Thomas Erlebach, Fabian Hennecke, Birgitta Weber, and Peter
  Widmayer.
\newblock Scheduling with release times and deadlines on a minimum number of
  machines.
\newblock In {\em IFIP TCS}, pages 209--222, 2004.

\bibitem{DevanurJK13}
Nikhil~R. Devanur, Kamal Jain, and Robert~D. Kleinberg.
\newblock Randomized primal-dual analysis of ranking for online bipartite
  matching.
\newblock In {\em SODA}, pages 101--107, 2013.

\bibitem{EnglertW12}
Matthias Englert and Matthias Westermann.
\newblock Considering suppressed packets improves buffer management in quality
  of service switches.
\newblock {\em SIAM J. Comput.}, 41(5):1166--1192, 2012.

\bibitem{GareyJ77}
M.~R. Garey and David~S. Johnson.
\newblock Two-processor scheduling with start-times and deadlines.
\newblock {\em SIAM J. Comput.}, 6(3):416--426, 1977.

\bibitem{GareyJ79}
Michael~R. Garey and David~S. Johnson.
\newblock {\em Computers and Intractability; A Guide to the Theory of
  NP-Completeness}.
\newblock W. H. Freeman \& Co., New York, NY, USA, 1979.

\bibitem{Hajek01}
Bruce Hajek.
\newblock On the competitiveness of on-line scheduling of unit-length packets
  with hard deadlines in slotted time.
\newblock In {\em CISS}, pages 434--439, 2001.

\bibitem{Jez11}
Lukasz Jez.
\newblock One to rule them all: A general randomized algorithm for buffer
  management with bounded delay.
\newblock In {\em ESA}, pages 239--250, 2011.

\bibitem{KaoCRW12}
Mong-Jen Kao, Jian-Jia Chen, Ignaz Rutter, and Dorothea Wagner.
\newblock Competitive design and analysis for machine-minimizing job scheduling
  problem.
\newblock In {\em ISAAC}, pages 75--84, 2012.

\bibitem{KesselmanLMPSS04}
Alexander Kesselman, Zvi Lotker, Yishay Mansour, Boaz Patt-Shamir, Baruch
  Schieber, and Maxim Sviridenko.
\newblock Buffer overflow management in qos switches.
\newblock {\em SIAM J. Comput.}, 33(3):563--583, 2004.

\bibitem{LiSS07}
Fei Li, Jay Sethuraman, and Clifford Stein.
\newblock Better online buffer management.
\newblock In {\em SODA}, pages 199--208, 2007.

\bibitem{RaghavanT87}
Prabhakar Raghavan and Clark~D. Thompson.
\newblock Randomized rounding: a technique for provably good algorithms and
  algorithmic proofs.
\newblock {\em Combinatorica}, 7(4):365--374, 1987.

\bibitem{Saha13}
Barna Saha.
\newblock Renting a cloud.
\newblock In {\em FSTTCS}, pages 437--448, 2013.

\bibitem{ShiY08}
Yongqiang Shi and Deshi Ye.
\newblock Online bin packing with arbitrary release times.
\newblock {\em Theor. Comput. Sci.}, 390(1):110--119, 2008.

\bibitem{Spieksma98}
Frits C.~R. Spieksma.
\newblock Approximating an interval scheduling problem.
\newblock In {\em APPROX}, pages 169--180, 1998.

\bibitem{Vazirani01}
V.~Vazirani.
\newblock {\em Approximation algorithms}.
\newblock Springer-Verlag, Berlin, 2001.

\bibitem{YuZ09}
Guosong Yu and Guochuan Zhang.
\newblock Scheduling with a minimum number of machines.
\newblock {\em Oper. Res. Lett.}, 37(2):97--101, 2009.

\end{thebibliography}

\end{document}